\documentclass[]{PoS}

\usepackage{amsmath}

\def\nslash{n\!\!\!\!\slash}
\def\bnslash{\bar n\!\!\!\!\slash}

\newcommand{\bn}{{\bar n}}

\newcommand{\vect}[1]{\mathbf{#1}}
\newcommand{\abs}[1]{\left\lvert #1\right\rvert}
\newcommand{\bra}[1]{\left\langle #1\right\rvert}
\newcommand{\ket}[1]{\left\lvert #1\right\rangle}

\newcommand{\Lqcd}{\Lambda_{\text{QCD}}}
\newcommand{\e}{\mathrm{e}}

\newcommand{\MeV}{\text{ MeV}}
\newcommand{\GeV}{\text{ GeV}}

\newcommand{\eq}[1]{Eq.~\eqref{#1}}
\newcommand{\eqs}[2]{Eqs.~\eqref{#1} and \eqref{#2}}

\DeclareMathOperator{\Tr}{Tr}

\newcommand{\as}{\alpha_s}

\title{Probing the Structure of Jets: Factorized and Resummed Angularity Distributions in SCET}

\ShortTitle{Angularity Distributions in SCET}

\author{\speaker{Christopher Lee}, Andrew Hornig, and Grigory Ovanesyan\\
        Department of Physics, University of California, and Theoretical Physics Division, Lawrence Berkeley National Laboratory, 
         Berkeley, CA  94720, USA\\
        E-mail: \email{clee@berkeley.edu, ahornig@berkeley.edu, ovanesyan@berkeley.edu}}

\abstract{Using the framework of soft-collinear effective theory (SCET), we factorize and calculate $e^+e^-$ angularity distributions, including perturbative resummation and the incorporation of a universal model for the nonperturbative soft function. Angularities $\tau_a$ are a class of event shapes varying in their sensitivity to the substructure of jets in the final state, controlled by a continuous parameter $a<2$. We calculate  the jet and soft functions in factorized angularity distributions for all $a<1$ to first order in the strong coupling $\alpha_s$ and resum large logarithms  to next-to-leading logarithmic (NLL) accuracy.   We employ a universal model for the nonperturbative soft function with a  gap parameter which cancels the renormalon ambiguity in the partonic soft function.}

\FullConference{International Workshop on Effective Field Theories: from the pion to the upsilon \\
		February 2-6 2009\\
		Valencia, Spain}

\begin{document}

\section{Introduction}

Event shapes yield simple information about the geometry of a hadronic final state in $e^+ e^-$ annihilation and can be used to probe the strong interactions at various energy scales \cite{Dasgupta:2003iq}. Two-jet event shapes $e$ are designed so that they take a numerical value, usually between 0 and 1, so that one of the kinematic endpoints (usually $e=0$) corresponds to events with two perfectly-collimated back-to-back jets in the final state. Event shape distributions depend on the hard-scattering cross-section at the large center-of-mass energy $Q$, on the perturbative branching and showering of the hard partons into jets at intermediate scales, and on the soft color exchange between jets and hadronization at a soft scale $\Lqcd$. Event shapes are thus are particularly useful probes of both perturbative and nonperturbative effects in QCD, allowing, for instance, extraction of the strong coupling $\as$ and nonperturbative shape function parameters  (see talks by I. Stewart and V. Mateu).

In this talk, I overview the calculation, performed in \cite{Hornig:2009vb}, of a particular class of event shape distributions, the \emph{angularities}, using the framework of soft-collinear effective theory (SCET) (see review talk by S. Fleming). We first establish a factorization theorem and calculate the hard, jet, and soft functions for angularity distributions to next-to-leading order (NLO) in $\as$. Then we solve the renormalization group equations for these functions and thereby resum large logarithms to next-to-leading logarithmic (NLL) accuracy in the angularity distributions near the two-jet kinematic endpoint. Finally we account for the effects of hadronization by introducing a universal model for the nonperturbative soft function, and present our final predictions for angularity distributions.

\section{Event Shapes and Angularities}

The most familiar  event shape is the thrust, $T = \frac{1}{Q}\max_{\vect{t}}\sum_{i\in X} \abs{\vect{t}\cdot\vect{p}_i}$, 
where $Q$ is the $e^+ e^-$ center-of-mass energy, and  $\vect{t}$, the \emph{thrust axis}, is the unit three-vector which maximizes the sum of projections of final-state particles' three-momenta onto this axis. Once the thrust axis is determined, many other event shapes can be defined, such as the jet broadening, $B = \frac{1}{Q}\sum_{i\in X} \abs{\vect{t}\times \vect{p}_i}$. 
A generalization of thrust and jet broadening is the class of \emph{angularities} \cite{Berger:2003iw},
\begin{equation}
\tau_a(X) = \frac{1}{Q}\sum_{i\in X}E_i \sin^a \theta_i (1-\cos\theta_i)^{1-a} = \frac{1}{Q}\sum_{i\in X} \abs{\vect{p}_i^T}\e^{-\abs{\eta_i}(1-a)}\,,
\end{equation}
where in the first form, $E_i$ is the energy of final-state particle $i$ and $\theta_i$ is its angle with respect to the thrust axis. In the second form, $\vect{p}_i^T$ is the $i$th particle's transverse momentum  and  $\eta_i$ its rapidity with respect to $\vect{t}$.  The parameter $a$ can be any real number, $-\infty<a<2$ for $\tau_a$ to be an infrared-safe observable. Two special cases are $a=0$ and $a=1$, which correspond to the thrust and jet broadening, $\tau_0 = 1 - T$ and $\tau_1 = B$.
It is known that the form of the factorization theorem which holds for the thrust distribution breaks down for the broadening distribution. By varying $a$ between 0 and 1, we can study how this factorization breaks down as $a$ approaches 1 \cite{Hornig:2009kv}. More generally, being able to vary $a$ provides a wealth of information about the final state that is otherwise obscured in looking at a single event shape in isolation. Roughly speaking, the angularity distributions for larger $a$ are dominated by jets which are very narrow, while those for smaller $a\to-\infty$ are dominated by jets which are wider.

\section{Energy Flow and Factorization in SCET}

The angularity distribution for events $e^+ e^-\to X$ at center-of-mass energy $Q$ is given by \cite{Bauer:2008dt}
\begin{equation}
\label{noX}
\frac{d\sigma}{d\tau_a} = \frac{1}{2Q^2} \int d^4 x \ \e^{iq\cdot x} \sum_{i=V,A} L_{\mu\nu}^i \bra{0} j_i^{\mu\dag}(x) \delta(\tau_a - \hat\tau_a) j_i^\nu(0)\ket{0}\,,
\end{equation}
where $q=(Q,\vect{0})$, $j_{V,A}^\mu = \bar q\Gamma_{V,A}^\mu q$ (with $\Gamma_V^\mu = \gamma^\mu$ and $\Gamma_A^\mu = \gamma^\mu\gamma_5$), and $L_{\mu\nu}^{V,A}$ are the leptonic tensors corresponding to these vector and axial currents (see \cite{Bauer:2008dt}). We have made use of an operator  $\hat\tau_a$ which acts on states $X$ and returns the eigenvalue $\tau_a(X)$.  This operator can be constructed from a momentum flow operator $\mathcal{E}_T(\eta)$, which in turn is constructed from the energy-momentum tensor \cite{Bauer:2008dt,Korchemsky:1997sy,Lee:2006nr},
\begin{equation}
\label{ETfromT0i}
\mathcal{E}_T(\eta) = \frac{1}{\cosh^3\eta}\int_0^{2\pi} d\phi \lim_{R\to \infty} R^2\int_0^\infty dt\ \hat n_i T_{0i}(t,R\hat n)\,,
\end{equation}
where $\hat n$ is a unit vector in the direction given by rapidity $\eta$ and azimuthal angle $\phi$. $R$ is the radius of a sphere centered on the $e^+ e^-$ collision.
$\mathcal{E}_T(\eta)$ acts on states $X$ according to $\mathcal{E}_T(\eta)\ket{X} = \sum_{i\in X}\abs{\vect{p}_i^T}\delta(\eta - \eta_i)\ket{X}$,
where the transverse momentum $\vect{p}_i^T$ and the rapidity $\eta_i$ are measured with respect to the thrust axis.\footnote{The operator thus implicitly depends on the thrust axis $\vect{t}$ of the final state. An operator which returns this axis was constructed in \cite{Bauer:2008dt} and can be used here. In SCET, however, no such operator is required, as the thrust axis can be identified (up to power corrections) with the collinear jet direction $\vect{n}$. This approximation breaks down for $a\geq 1$, which is one manifestation of the breakdown of factorization for these values of $a$ \cite{Berger:2003iw,Bauer:2008dt}.} 
In terms of $\mathcal{E}_T(\eta)$, the operator $\hat \tau_a$ can be constructed,
\begin{equation}
\hat\tau_a = \frac{1}{Q}\int_{-\infty}^\infty d\eta\ \e^{-\abs{\eta}(1-a)} \mathcal{E}_T(\eta)\,.
\end{equation}

To factorize and evaluate the distribution \eq{noX} in SCET, we match the QCD current $j_i^\mu$ and the operator $\hat\tau_a$ onto operators in SCET. The current matches onto two-jet operators in SCET containing collinear fields in two back-to-back directions \cite{Bauer:2002nz,Bauer:2003di},
\begin{equation}
j_i^\mu(x) = \sum_{\vect{n}}\sum_{\tilde p_n,\tilde p_\bn} \e^{i(\tilde p_n - \tilde p_\bn)\cdot x} C_{n\bn}(\tilde p_n,\tilde p_\bn;\mu) \bar\chi_{n,p_n}(x) Y_n(x) \Gamma_i^\mu \overline Y_{\bn}(x)\chi_{\bn, p_\bn}(x)\,,
\end{equation}
where $\Gamma_V^\mu = \gamma_\perp^\mu, \Gamma_A^\mu = \gamma_{\perp}^\mu \gamma_5$, and the sums are over the direction $\vect{n}$ of the light-cone vectors $n,\bn = (1,\pm\vect{n})$, and the label momenta $\tilde p_{n,\bn}$ \cite{Bauer:2000yr,Bauer:2001ct}. The jet fields $\chi_{n,\bn} \equiv W_{n,\bn}^\dag \xi_{n,\bn}$ are built up from collinear quark fields $\xi_{n,\bn}$ and collinear Wilson lines  $W_{n,\bn}$, which have been decoupled from soft gluons through the BPS field redefinition with soft Wilson lines $Y_n,\overline Y_\bn$ \cite{Bauer:2001yt}. The Wilson lines are defined in \cite{Bauer:2000yr,Bauer:2001ct}.  The matching coefficient $C_{n\bn}$ is calculable in perturbation theory.

To match $\hat \tau_a$ onto SCET operators, we simply 
replace the energy-momentum tensor $T_{0i}$ in \eq{ETfromT0i} with the energy-momentum tensor in SCET. Since the Lagrangian of SCET (after the BPS field redefinition) splits into separate purely collinear and purely soft parts, so does the energy-momentum tensor, and, thus, also the event shape operator. That is, $\hat\tau_a = \hat\tau_a^n + \hat\tau_a^\bn + \hat\tau_a^s$, 
where $\hat \tau_a^{n,\bn,s}$ are constructed as above, but using the energy-momentum tensor of only the $n,\bn$-collinear or soft Lagrangian of SCET. We can then factorize the angularity distributions \eq{noX} in SCET,
\begin{equation}
\label{factorization}
\frac{1}{\sigma_0} \frac{d\sigma}{d\tau_a} = H(Q;\mu) \int d\tau_a^n\ d\tau_a^\bn\ d\tau_a^s \delta(\tau_a - \tau_a^n - \tau_a^\bn - \tau_a^s) J_a^n(\tau_a^n;\mu) J_a^\bn(\tau_a^\bn;\mu) S_a(\tau_a^s;\mu)\,,
\end{equation}
where $\sigma_0$ is the total $e^+ e^-\to q\bar q$ Born cross-section, the hard coefficient $H$ is the squared amplitude of the two-jet matching coefficients, $H(Q;\mu) = \abs{C_{n\bn}(Qn/2,-Q\bn/2;\mu)}^2$, 
and the jet and soft functions $J_a^{n,\bn},S_a$ are given by matrix elements of collinear and soft operators,
\begin{align}
J_a^n(\tau_a^n;\mu) \left(\frac{\nslash}{2}\right)_{\alpha\beta} &= \frac{1}{N_C}\int \frac{dl^+}{2\pi}\int d^4 x\ \e^{il\cdot x} \Tr \bra{0}\chi_{n,Q}(x)_\alpha \delta(\tau_a^n - \hat\tau_a^n)\bar\chi_{n,Q}(0)_\beta\ket{0} \\
J_a^\bn(\tau_a^\bn;\mu) \left(\frac{\bnslash}{2}\right)_{\alpha\beta} &= \frac{1}{N_C}\int \frac{dk^-}{2\pi}\int d^4 x\ \e^{ik\cdot x} \Tr \bra{0}\bar\chi_{\bn,-Q}(x)_\beta \delta(\tau_a^\bn - \hat\tau_a^\bn)\chi_{\bn,-Q}(0)_\alpha\ket{0} \\
S_a(\tau_a^s;\mu) &= \frac{1}{N_C}\Tr\bra{0} \overline Y_\bn^\dag(0) Y_n^\dag(0)\delta(\tau_a^s - \hat\tau_a^s) Y_n(0) \overline Y_\bn(0) \ket{0}\,,
\end{align}
where the traces are over colors.

\section{Renormalization Group Evolution and Resummation}

A fixed-order calculation of the angularity distributions will be divergent in the endpoint region $\tau_a\to 0$ where the event becomes more and more two-jet-like. The divergent terms $\sim \as^n(\ln^m\tau_a)/\tau_a$ must be resummed to all orders in $\as$ to yield a reliable, finite prediction. This resummation can be achieved by solving renormalization group equations for the hard, jet, and soft functions defined in the factorized angularity distributions above, and running each function from the scale at which large logarithms in it are minimized to the common factorization scale $\mu$ \cite{Contopanagos:1996nh}.

We begin by calculating the hard, jet, and soft functions  to fixed order in perturbation theory using the Feynman rules of SCET \cite{Bauer:2000yr,Bauer:2001yt}. The result for the renormalized hard function is \cite{Bauer:2003di,Manohar:2003vb}
\begin{equation}
H(Q;\mu) = 1 - \frac{\as C_F}{2\pi}\left(8 - \frac{7\pi^2}{6} + \ln^2\frac{\mu^2}{Q^2} + 3\ln\frac{\mu^2}{Q^2}\right)\,,
\end{equation}
the renormalized jet function is \cite{Hornig:2009vb}
\begin{align}
\label{jetNLO}
J_a^n (\tau_a^n; \mu) =  \delta(\tau_a^n) \bigg \{ 1 &+ \frac{\as C_F}{\pi} \bigg [  \frac{1-a/2}{2(1-a)} \ln^2{\frac{\mu^2}{Q^2}} + \frac{3}{4} \ln{\frac{\mu^2}{Q^2}} + f(a) \bigg] \bigg \}\nonumber\\
&  - \frac{\as C_F}{\pi} \left [ \bigg( \frac{3}{4} \frac{1}{1-a/2}+\frac{2}{1-a}\ln{\frac{\mu}{Q (\tau_a^n)^{1/(2-a)}}}\bigg )\bigg ( \frac{\theta(\tau_a^n)}{\tau_a^n}\bigg )\right ]_+ 
\,,\end{align}
where we defined
\begin{align}
\label{fa}
f(a) \equiv & \,\frac{1}{1-\frac{a}{2}}\bigg ( \frac{7-\frac{13a}{2}}{4} - \frac{\pi^2}{12}\frac{3 - 5a + \frac{9a^2}{4}}{1-a}  -\int_0^1\! d x \frac{1-x + \frac{x^2}{2}}{x} \ln[(1-x)^{1-a}+x^{1-a}] \bigg )
\,,\end{align}
and the renormalized soft function is  \cite{Hornig:2009vb}
\begin{align}
S_a^{\text{PT}}(\tau_a^s; \mu)  &= \delta(\tau_a^s) \left [1- \frac{\as C_F}{ \pi (1-a)}\left ( \frac{1}{2}\ln^2{\frac{\mu^2}{Q^2}} -\frac{\pi^2}{12} \right ) \right ] +\frac{2 \as C_F}{ \pi (1-a)} \left [ \frac{\theta(\tau_a^s)}{\tau_a^s} \ln{\frac{\mu^2}{(Q \tau_a^s)^2}}\right ]_+  
\label{softNLO}
\,.\end{align}
The superscript ``PT'' denotes that this is only the partonic soft function, calculable in perturbation theory. Later, we will convolute it with a model for the nonperturbative soft function. The perturbative calculations leading to these results fail to remain infrared-safe for $a\geq 1$, manifesting the breakdown of factorization for these values of $a$ \cite{Hornig:2009kv}.

The hard function obeys the renormalization group (RG) equation
\begin{equation}
\label{hardRGE}
\mu\frac{d}{d\mu} H(Q;\mu) = \gamma_H(\mu) H(Q;\mu)\,,
\end{equation}
where the anomalous dimension $\gamma_H$ takes the form
\begin{equation}
\gamma_H(\mu) = \Gamma_H[\as] \ln\frac{\mu^2}{Q^2} + \gamma_H[\as]\,.
\end{equation}
To first order in $\as$, $\Gamma_H[\as] = -2\as C_F/\pi$ and $\gamma_H[\as] = -3\as C_F/\pi$.
Meanwhile, the jet and soft functions obey the slightly more complicated RG equation,
\begin{equation}
\label{FRGE}
\mu \frac{d}{d\mu}F(\tau;\mu) = \int_{-\infty}^\infty d\tau' \ \gamma_F(\tau-\tau';\mu) F(\tau';\mu)\,,
\end{equation}
where $F=J, S$. The anomalous dimensions take the form
\begin{equation}
\gamma_F(\tau-\tau';\mu) = -\Gamma_F[\as]\left\{\frac{2}{j_F}\left[\frac{\theta(\tau-\tau')}{\tau-\tau'}\right]_+ - \ln\frac{\mu^2}{Q^2}\delta(\tau-\tau')\right\} + \gamma_F[\alpha_s]\delta(\tau-\tau')\,,
\end{equation}
where $j_J = 2-a$ and $j_S = 1$, and to first order in $\as$, 
\begin{equation}
\Gamma_J[\as] = \frac{2\as C_F}{\pi}\frac{1-a/2}{1-a} \,,\quad \Gamma_S[\as] = -\frac{2\as C_F}{\pi}\frac{1}{1-a}\,,\quad \gamma_J[\as] = \frac{3\as C_F}{2\pi} \,,\quad \gamma_S[\as] = 0 .
\end{equation}

The solution to the RG equation \eq{hardRGE} for the hard function is 
\begin{equation}
\label{hardsol}
H(Q;\mu) = H(Q;\mu_0) e^{K_H(\mu,\mu_0)}\left(\frac{\mu_0}{Q}\right)^{\omega_H(\mu,\mu_0)}\,,
\end{equation}
and to the RG equation \eq{FRGE} for the jet and soft functions,
\begin{equation}
\label{Fsol}
F(\tau;\mu) = \int d\tau' U_F(\tau-\tau';\mu,\mu_0) F(\tau';\mu_0)\,,
\end{equation}
where the evolution kernel $U_F$ is given by \cite{Fleming:2007xt}
\begin{equation}
\label{Fkernel}
U_F(\tau - \tau';\mu,\mu_0) = \frac{e^{K_F + \gamma_E\omega_F}}{\Gamma(-\omega_F)} \left(\frac{\mu_0}{Q}\right)^{j_F\omega_F} \left[\frac{\theta(\tau-\tau')}{(\tau-\tau')^{1+\omega_F}}\right]_+\,.
\end{equation}
In \eqs{hardsol}{Fkernel}, 
\begin{align}
\omega_F(\mu,\mu_0) &=  \frac{2}{j_F}\int_{\as(\mu_0)}^{\as(\mu)}\frac{d\alpha}{\beta[\alpha]}\Gamma_F[\alpha]\,,\\
K_F(\mu,\mu_0) &= \int_{\as(\mu_0)}^{\as(\mu)}\frac{d\alpha}{\beta[\alpha]}\gamma_F[\alpha] + 2\int_{\as(\mu_0)}^{\as(\mu)}\frac{d\alpha}{\beta[\alpha]}\Gamma_F[\alpha]\int_{\alpha_s(\mu_0)}^\alpha\frac{d\alpha'}{\beta[\alpha']}\,,
\end{align}
where $\beta[\alpha]$ is the beta function of QCD, and now $F= H,J,S$, with $j_H = 1$. 

The solutions \eqs{hardsol}{Fsol} for the hard, jet, and soft functions can be used to express these functions at the factorization scale $\mu$ in \eq{factorization} in terms of their values at (arbitrary) scales $\mu_{H,J,S}$. These scales can be chosen to minimize the logarithms in the fixed-order hard, jet, and soft functions, and then the RG running of each function to the scale $\mu$ resums logarithms of $\mu/\mu_{H,J,S}$. After convoluting the functions to obtain the full distribution, we find that the initial scales should be chosen near $\mu_H \sim Q$, $\mu_J\sim Q\tau_a^{1/(2-a)}$, and $\mu_S\sim Q\tau_a$. The running to $\mu$ (typically chosen near $\mu\sim Q$) achieves resummation of the $\log^m\tau_a/\tau_a$ divergent terms in the fixed-order angularity distributions. In the following results, we used the anomalous dimension $\Gamma_{H,J,S}[\as]$ to two loops and $\gamma_{H,J,S}[\as]$ to one loop to achieve NLL accuracy in the resummation of these terms.

\section{Model of Nonperturbative Effects}

A reliable prediction of the angularity distributions requires consistent incorporation of the above perturbative calculations and a universal model for the nonperturbative soft function. Such a model which interpolates consistently (i.e. with well-defined RG evolution) between perturbative and nonperturbative regimes is the convolution \cite{Hoang:2007vb,Ligeti:2008ac},
\begin{equation}
\label{softmodel}
S_a(\tau_a;\mu) = \int d\tau_a' S_a^{\text{PT}}(\tau_a - \tau_a';\mu) f_a^{\text{exp}}\left(\tau_a'  - \frac{2\Delta_a}{Q}\right) \,,
\end{equation}
where $f_a^{\text{exp}}$ is a model function based on that introduced for thrust and other event shape distributions in \cite{Korchemsky:2000kp}, which we have generalized here to arbitrary $\tau_a$ \cite{Hornig:2009vb}. $\Delta_a$ is a ``gap parameter'' \cite{Hoang:2007vb} that accounts for the minimum value of $\tau_a$ due to hadronization in each hemisphere. Specifically,
\begin{equation}
\label{fexp}
f_a^{\text{exp}} (\tau_a) = \theta(\tau_a) \mathcal{N}(A,B) \frac{Q}{\Lambda_a}\left(\frac{Q\tau_a}{\Lambda_a}\right)^{\!\!\! 2A-1}\!\! {}_1 F_1\left(\frac{1}{2},\frac{1}{2}+A,\frac{B-1}{2}\left(\frac{Q\tau_a}{\Lambda_a}\right)^{\!\!\!2}\right)\exp\left[-\frac{B+1}{2}\left(\frac{Q\tau_a}{\Lambda_a}\right)^{\!\!\!2}\right]\,,
\end{equation}
where $\Delta_a = \Delta/(1-a)$ and $\Lambda_a = \Lambda/(1-a)$ are related to the $a=0$ (thrust) model parameters. We choose typical values for the model parameters (cf.~\cite{Korchemsky:2000kp}), $A = 2.5$, $B = -0.4$, and $\Lambda = 0.55\GeV$. The constant $\mathcal{N}(A,B)$ normalizes the total integral of $f_a^{\text{exp}}$  to 1. \eq{fexp} is the simplest generalization of the thrust model function that obeys the universal scaling of the first moment of the soft function, $\langle \tau_a\rangle_S \sim 1/(1-a)$, proposed in \cite{Berger:2003pk} and proven to all orders in $\as$ in \cite{Lee:2006nr}.

The gap parameter $\Delta_a$ and partonic soft function $S_a^{\text{PT}}$ contain  $1/Q$ renormalon ambiguities, which can be removed by shifting the gap by a perturbatively-calculable amount, $\Delta_a = \bar\Delta_a(\mu) + \delta_a(\mu)$ where $\bar\Delta_a$ is renormalon-free, and we choose the perturbative shift $\delta_a$ to be\begin{equation}
\delta_a(\mu) = -\frac{Q}{2}\frac{\int d\tau_a\, \tau_a \e^{-Q\tau_a/(R\e^{\gamma_E})}S_a^{\text{PT}}(\tau_a;\mu)}{\int d\tau_a\, \e^{-Q\tau_a/(R\e^{\gamma_E})}S_a^{\text{PT}}(\tau_a;\mu)} = -R\e^{-\gamma_E}\frac{8C_F}{1-a}\frac{\as(\mu)}{4\pi}\ln\frac{\mu}{R} + \mathcal{O}(\as^2)\,,
\end{equation}
where different choices of the free parameter $R$ determine different renormalon subtraction schemes  \cite{Jain:2008gb,Hoang:2008fs}. Expanding the soft function \eq{softmodel} in powers of $\as$, $S_a^{\text{PT}}$ combines with $\delta_a$-dependent terms to become renormalon-free, as is the new gap parameter $\bar\Delta_a$. These ``$R$'' schemes provide a consistent evolution equation for $\Delta_a$ in both $\mu$ and $R$ (see talk by I. Scimemi). Below, we choose the renormalon-free gap to be $\Delta_a(\mu=1\GeV) = 100\MeV$ and $R = 100\MeV$. 

\section{Final Predictions for Angularity Distributions}

\begin{figure}
\vspace{-1ex}
\centerline{\includegraphics[width=.63\textwidth]{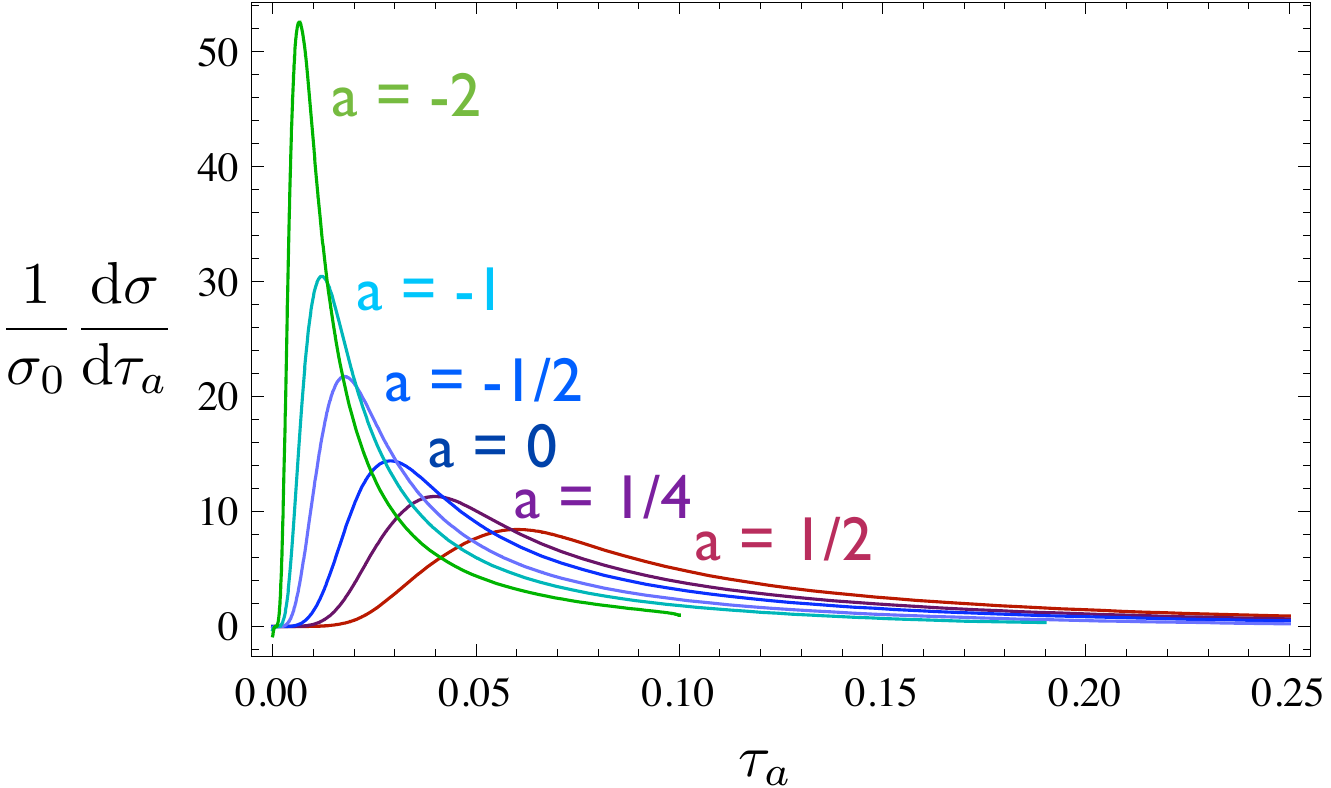}}
\vspace{-1em}
\caption{Angularity distributions for $-2<a<\frac{1}{2}$ at $Q= 100\GeV$, with $\mathcal{O}(\as)$ hard, jet, and soft functions, NLL resummation, and gapped model soft function.}
\label{fig1}
\end{figure}

\begin{figure}
\vspace{-.5ex}
\centerline{\includegraphics[width=\textwidth]{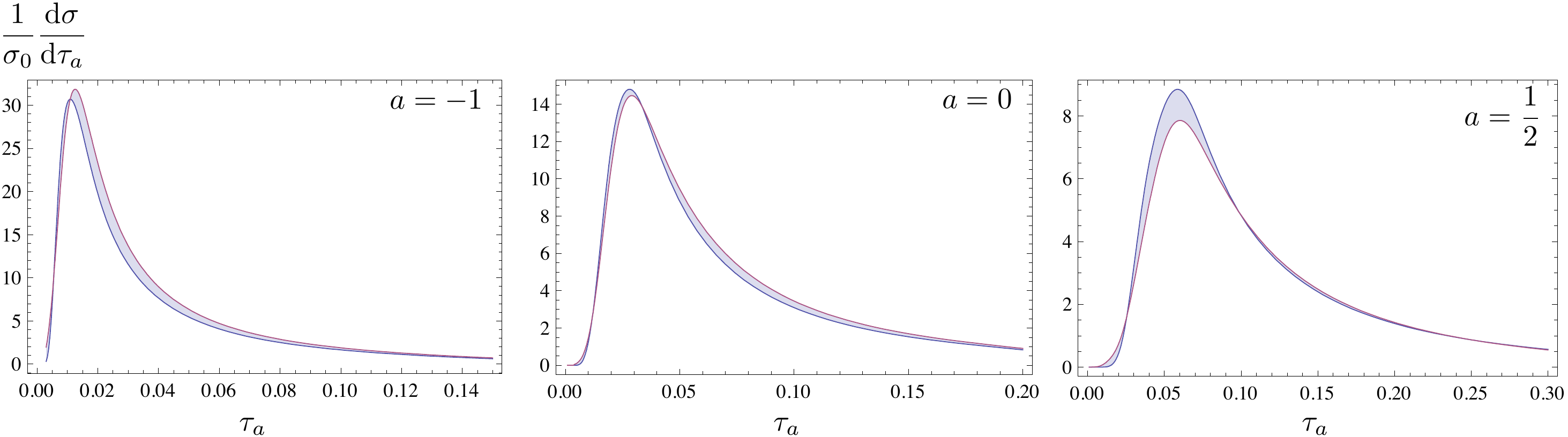}}
\vspace{-1em}
\caption{Correlated variation of jet and soft scales $\mu_J$ and $\mu_S$ together by factors of $1/2$ and $2$ in angularity distributions at $Q = 100\GeV$ for $a=-1$, $a=0$, and $a=\frac{1}{2}$. }
\label{fig2}
\end{figure}

In Fig.~\ref{fig1} we plot  angularity distributions for $-2 \leq a \leq1/2$ for $Q = 100\GeV$, including the above $\mathcal{O}(\as)$ fixed-order calculations of the hard, jet, and soft functions, the NLL resummation in the $\tau_a\to 0$ endpoint region to all orders in $\as$, and  the soft model function of \eq{fexp}. We also matched the resummed distribution, which is accurate in the endpoint region, to fixed-order QCD, which is more accurate in the tail region, using the procedure described in \cite{Hornig:2009vb}. We choose the hard scale to be $\mu_H = Q$, and the jet and soft scales to interpolate between a minimum fixed value $\mu_{J,S} = \mu_{J,S}^{\text{min}}\gtrsim\Lqcd$ for small $\tau_a$ and $\mu_J = Q\tau_a^{1/(2-a)}$ and $\mu_S = Q\tau_a$ for larger $\tau_a$, allowing us to avoid spurious Landau poles \cite{Hornig:2009vb}. In the plots, we chose $\mu_S^{\text{min}} = 1\GeV$ and $\mu_J^{\text{min}} = Q^{(1-a)/(2-a)}(\mu_S^{\text{min}})^{1/(2-a)}$. 
In Fig.~\ref{fig2} we plot the $a=-1, 0, 1/2$ distributions varying the scales  $\mu_{J,S}$ together (correlated) by factors of 2. Dependence on $\mu$ and $\mu_H$ is much weaker.

\section{Conclusion}

We have predicted angularity distributions resummed in perturbation theory to NLL accuracy, including for the first time the jet and soft functions in the factorization theorem to NLO, and a universal model for the nonperturbative soft function. We used the framework of SCET to perform the factorization, resummation, and incorporation of the nonperturbative model  in a unified way. Comparison to data from LEP or a future linear collider will test the robustness of the model we employed for the nonperturbative soft function. Extension of the notion of event shapes to individual ``jet shapes'' can also allow the probing of jet substructure in hadron collisions \cite{Almeida:2008yp,Almeida:2008tp}.

This work was supported in part by the U.S. Department of Energy under Contract DE-AC02-05CH11231 and the National Science Foundation under grant PHY-0457315.

\bibliography{NLL}

\end{document}